# Achieving High-Performance Fault-Tolerant Routing in HyperX Interconnection Networks

Cristóbal Camarero*, Alejandro Cano*, Carmen Martínez*, Ramón Beivide*†

{cristobal.camarero, alejandro.cano, carmen.martinez, ramon.beivide}@unican.es

*Universidad de Cantabria, SPAIN

†The contributions from R. Beivide were made, in part, while he was affiliated with Barcelona Supercomputing Center.

## Abstract

Interconnection networks are key actors that condition the performance of current large datacenter and supercomputer systems. Both topology and routing are critical aspects that must be carefully considered for a competitive system network design. Moreover, when daily failures are expected, this tandem should exhibit resilience and robustness. Low-diameter networks, including HyperX, are cheaper than typical Fat Trees. But, to be really competitive, they have to employ evolved routing algorithms to both balance traffic and tolerate failures.

In this paper, SurePath, an efficient fault-tolerant routing mechanism for HyperX topology is introduced and evaluated. SurePath leverages routes provided by standard routing algorithms and a deadlock avoidance mechanism based on an Up/Down escape subnetwork. This mechanism not only prevents deadlock but also allows for a fault-tolerant solution for these networks. SurePath is thoroughly evaluated in the paper under different traffic patterns, showing no performance degradation under extremely faulty scenarios.

## 1 Introduction

High-performance computing (HPC) is based on parallelism. The practical exploitation of such parallelism depends, up to some extent, on the interconnection network present in any HPC computing platform. Nowadays, from Networks on Chip (NoCs) to big datacenter system networks, the role of these communication subsystems is considered critical in computer architecture and engineering. The emphasis of this paper is put on system networks for large datacenters and supercomputers.

The Folded Clos topology [9], or Fat Tree, has been used in many of these large systems. In such networks, only the leaf layer of switches connects servers; whereas the remainder of the switches, organized in further layers, employ their ports just to connect other switches. Currently, large computer systems begin to use other kind of topologies in which all their switches connect both servers and other switches. Indeed, the advent of large degree switches has enabled the possibility of using Complete graphs, in which every pair of vertices is connected by a link, as the topology for interconnection networks with a moderate number of servers. For example, with 64-port switches, one can build a network with 33 switches, based on the Complete graph $K_{33}$, equipping 1056 servers (32 servers per switch) and employing 528 wires. Larger deployments require either hierarchical or multi-dimensional topologies based on Complete graphs, such as Dragonflies [20] and Flattened Butterflies [19] or HyperX networks [1]. Typical Dragonflies are based on Complete graphs both for the local topologies of the groups and for the global network. The Dragonfly+ network [33], is similar to a Dragonfly but uses leaf-spine Fat Tree groups instead of Complete graphs. HyperX networks are Cartesian graph products of Complete graphs, which in Graph Theory are denoted as Hamming graphs. These new topologies are significantly cheaper than Fat Trees, which has increased the interest on them [17].

In large HPC or datacenter systems, a few daily failures are expected [26]. Link or switch failures are easy to manage when using Fat Trees due to their redundancy but such a redundancy entails a high cost. For example in the Blue Gene [13], cables between racks have spare wires that can be used to replace faulty links, so the actual topology is not modified. However, this can be prohibitive for networks with lots of expensive cabling. In other cases, such as Tofu, the presence of a fault forces a reconfiguration that disconnect additional nodes to keep the logical topology mostly intact, sacrificing computational resources [2]. Other proposed strategies sacrifice performance to maintain the service [29].

Hamming graphs, which are the base of HyperX, have a very rich structure, and are around a 25% cheaper than Fat Trees. However, a single failure can make several specifically devised routing algorithms to fail. Very general routing algorithms, such as Minimal, keep working, only requiring to run a BFS to recompute the routing tables. However, more specialised algorithms such as Dimension Ordered Routing (DOR) and OmniWAR [24] cannot work if a single link is removed. In the case of DAL, the routing originally proposed for HyperX networks, only supports one fault in the network. There-fore, we tackle the problem by proposing a mechanism that works for HyperX networks.

This work proposes *SurePath*, a routing mechanism for HyperX interconnection networks that allows for fault-tolerance requiring minimal resources. This routing mechanism uses a set of routes provided by a routing algorithm such as Omnidimensional [24] or Polarized [5], together with a novel escape subnetwork. These base routings are progressively adaptive non-minimal routings, with Omnidimensional assuming a HyperX topology and Polarized discovering the topology (routing tables filled by a Breadth-First Search) at boot time, upgrade or failure. The escape subnetwork follows an adaptive routing mechanism which is deadlock-free only with one buffer FIFO per port. It serves for both, deadlock avoidance and fault tolerance, thus allowing the remaining virtual channels (VCs for short) to be used for performance instead of being dedicated to avoid deadlock. This escape subnetwork employs routing tables, which can be updated when a failure occurs. The whole mechanism is guaranteed to work while there are possible paths and the throughput degrades gracefully with the amount of faults.

The main contributions of this paper are the following:

1. The proposal of SurePath, a new routing mechanism for HyperX networks that tolerates faults at a low cost and high performance.

2. An exhaustive empirical evaluation, based on simulation, measuring the performance of the proposed solution on HyperX networks, with two routing algorithms: Omnidimensional [24] and Polarized [5].

3. An evaluation of the proposed mechanism over a HyperX in different faulty scenarios, showing its immunity to the presence of a great number of random faults.

4. A new traffic pattern for which Omnidimensional cannot reach optimal performance, which helps to analyze behaviours depending on using a different set of base routes.

The remainder of the paper is organized as follows. Section 2 motivates the proposal of SurePath routing mechanism, which is described in Section 3. In Section 4 the experimental set-up is described. In Section 5 SurePath routing mechanism is evaluated and compared to other routings in a fault free scenario. Next, Section 6 shows that SurePath constitutes a good fault-tolerant routing solution for HyperX networks by evaluating its performance in different faulty scenarios. Finally, in Section 7 the main findings of the paper are dicussed.

## 2 Motivation

This work is based on the idea that, for being competitive against the Fat Tree, a modern low-diameter network must be equipped with a practicable routing mechanism, able to tolerate failures. With this aim, we have designed SurePath, a robust low-cost efficient routing mechanism able to deal with a high number of failures in HyperX networks.

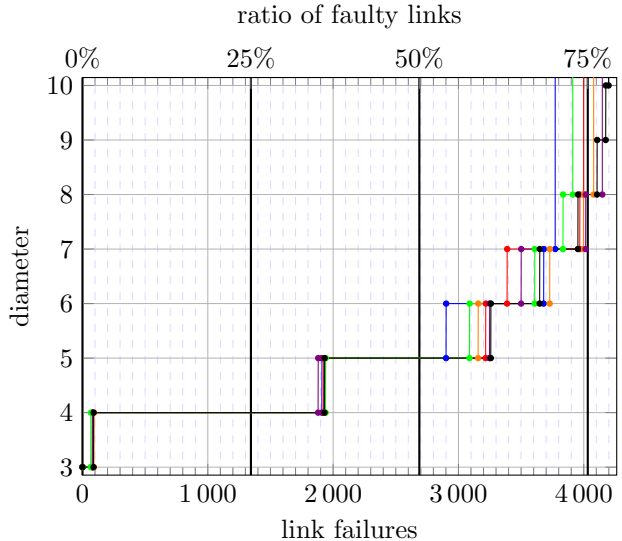

Figure 1: Evolution of distances in a $8 \times 8 \times 8$ Hyper X with increasing random uniform failures. The network becomes disconnected as the line exits the plot.

A $nD$ HyperX is defined for sides $k_1 \times \cdots \times k_n$ as a network topology with $\prod_{i=1}^{n} k_i$ switches. Typically, $k_1 = k_2 = \cdots = k_n = k$ and each switch is connected to $k$ servers. This results in $k^n$ switches and $k^{n+1}$ servers. Some definitions include the possibility of parallel links between two switches or servers, but we do not consider it. For practical purposes, the dimension $n$ is 2 or 3. A switch $x$ is labelled by the vector of its coordinates $(x_1, \ldots, x_n)$ and it is connected to each other switch $y$ with coordinates $(y_1, \ldots, y_n)$ in which $x_i = y_i$ in all but one coordinate. Equivalently, the switch $x$ is connected with $y$ when their Hamming distance is 1. As the graph distance coincides with the Hamming distance, this topology has been called *Hamming graph* in some forums. The HyperX can also be represented as the Cartesian product of Complete graphs, which for the case of having the same number of switches per dimension, can be written $K_k^n$. With this perspective, each Complete subgraph $K_k$ is a row or column of switches, in which all pairs of switches are connected.

The resiliency properties of HyperX networks have been known for a long time [22]. For example, worst case faults are determined in [22, Corollary 5.2] and more recently the number of paths under failures is calculated in [30]. However, let us illustrate the robustness of HyperX with a quick look to Figure 1. The figure shows how HyperX diameter increases as the number of failures of randomly located links grows. Each colour represents a different sequence of random faults. It represents the diameter versus the number of random failures in a $8 \times 8 \times 8$ 3D HyperX. It should be highlighted that about 80 random link failures are needed to increment the diameter from 3 to 4. About 35% of the links must fail to increase the diameter to 5, and about 75% to disconnect the network.

It is not trivial to take advantage of HyperX's resiliency. For example, for each pair source/destination, DOR gives a unique route. If any link in this path fails, then packets for that pair cannot be delivered. The situation with OmniWAR is slightly better. Consider the situation where one link has failed and the traffic pattern requires to misroute, like Dimension Complement Reverse from [24]. Then, for the two endpoints of the faulty link, about $\frac{1}{n(k-1)}$ of the traffic would come from a route finishing on the faulty link. In such traffic, most deroutes are consumed at the start of the route and upon arriving to the fault link there is no longer possibility for it.

These routings could be augmented with a escape subnetwork, through which to send these packets. This escape subnetwork would be built considering the faults, to guarantee a path to the destination. However, this imposes extra load for a subnetwork supposedly dedicated to prevent deadlock. Other routings that may take into account changes to the topology can better adapt to faults, introducing less load into the escape subnetwork.

SurePath uses an escape subnetwork for deadlock avoidance. This mechanism will be used with Omnidimensional routing and Polarized routing. Omnidimensional is considered since it is the most accepted routing for HyperX networks. Polarized routing has been selected because, in a fault-free scenario, it achieves better performance than other competitors with similar costs. In both cases, the use of SurePath instead of their standard deadlock avoidance mechanisms provides high performance even under faulty-scenarios.

It has been argued that using escape paths has major limitations for large-scale networks [24]. But this depends on several factors such as the complexity of the employed flow-control and the router implementation. The authors of [24] assert that the only way to practically implement escape paths in a high-radix router is to use atomic queue allocation, reducing channel efficiency to one packet per VC. However, there are techniques that avoid such a stringent condition by using Virtual Cut-through flow control, such as the one presented in [23]. With respect to switch implementation, it is clear that some hierarchical switch architectures present severe difficulties in employing escape channels. This is the case, for example, of the routers proposed in [21] and [32], since they use Wormhole flow-control inside their data-paths. Nevertheless, escape channels have been used with no difficulty in other hierarchical switches. In Tianhe-2 [10], for example, its high-degree switches employ Virtual Cut-though inside the switch ASIC, that allows for using *Bubble* escape channels [6], as proved and implemented in that work. Another particular implementation would be the IBM's SCOC switch [8], which is bufferless, so this restriction on using escape channels in high-degree switches does not apply.

## 3 SurePath Routing Mechanism

In SurePath routing mechanism the virtual channels are separated into two sets, denoted as $C_{Rout}$ and $C_{Esc}$.

The set $C_{Rout}$ is used as the subnetwork responsible to accept the bulk of load, by means of a fully-adaptive routing algorithm. The two routings considered for this subnetwork in the paper are explained later in Subsection 3.1.

The set $C_{Esc}$ is used to implement the escape subnetwork as its deadlock avoidance mechanism, as defined, for example, in [12, 27]. Using such a policy, not only efficiently avoids packet deadlocks, but also achieves high performance. Moreover, it reduces the number of VCs required and allows for tolerating failures in the network. In SurePath, the escape subnetwork consists of a set of Up/Down routes plus opportunistic horizontal links, or shortcuts, that guarantees deadlock-freedom. The escape subnetwork is intended to be used as the last resource. Deadlock is globally avoided by allowing every packet in a channel $C_{Rout}$ to request a channel in the escape subnetwork $C_{Esc}$. Finally, in SurePath it is not allowed to move from a escape channel into a routing one.

The construction of both subnetworks ensures they allow a table-based implementation in which the current router may employ an internal table indexed with source and/or destination tag to decide the valid ports for the next hop and give preferences to them. Furthermore, these tables can be computed by a BFS algorithm when the topology changes, which keeps cost in the order of using Minimal routing.

The rules of the possible transitions between the two sets of VCs, $C_{Rout}$ and $C_{Esc}$ in SurePath, are given next, each case giving a penalization in phits $P$.

1. Every packet in $C_{Rout}$ has as candidates for the next hop the neighbours returned by the routing algorithm. The offered legal VCs have a penalty depending on their quality, ordered from greater to lowest.

2. Every packet, regardless of being on $C_{Rout}$ or $C_{Esc}$ has all neighbours in the escape subnetwork as candidates using $C_{Esc}$. Up/Down channels are highly penalized to avoid congestion on the root. Shortcuts are penalized depending on their reduction on the Up/Down distance.

Naturally, the hops that enter in $C_{Rout}$, using item 1), are called *routing hops*. The other hops, which enter a channel in $C_{Esc}$ by item 2), are called *escape hops*. Forced hops are just the cases in which packets in $C_{Rout}$, in item 1), do not obtain any routing candidate, so the neighbour must be taken from the escape subnetwork (item 2).

The penalties $P$ of the allowed neighbour ports are combined with their queue occupancies $Q$ to determine the output channel request. This is done by considering $Q$ as $q_s + \sum_{q \in \text{port}(q_s)} q$, where $q_s$ is the occupancy of the buffer to be requested and $\text{port}(q_s)$ is the set of all queues in the same port as $q_s$. Note that these values $q_s$ are calculated by adding together the output buffer occupancy and the count of consumed credits for the neighbour router. In other words, all queues of the requested port are added, counting twice the specific queue being requested. Then, the packet makes a single request to the output with the lowest $Q + P$ if it satisfies flow-control requirements. Ties are resolved randomly.

The specific values of $P$ have been chosen experimentally to give good performance. The ideal case may depend on architectural aspects such as the FIFO queue length and implementation details. Nevertheless, there are large regions of similar performance, so the specific values have little importance.

The following subsections give the details of the routings employed in each of the subnetworks.

## 3.1 Routing Algorithms in HyperX

### 3.1.1 Omnidimensional Routes

Omnidimensional routing algorithm was utilized in DAL [1] and OmniWAR routings [24].

At each hop in Omnidimensional routing, a packet may only move through any dimension in which the coordinates of the current switch differ from the destination switch. And for each of these dimensions, all the neighbour routers through that unaligned dimension are valid candidates for the next hop. These candidates include all the possible minimal hops and many potential deroutes. A limit of $m$ non-minimal hops is imposed as parameter, forcing further hops to be minimal. This limit is global, specifically allowing more than one deroute in the same dimension. Thus, it allows each packet to perform up to $n + m$ hops, where $m$ is the non-minimal hops permitted, and $n$ is the number of dimensions in the HyperX. For some traffic patterns it is necessary to have $m \geq n$ to have acceptable performance, and $m = n$ is always enough, which we assume it henceforth.

The candidates in minimal routes are given priority when making allocator requests. Specifically, we set $P = 0$ phits for minimal candidates and $P = 64$ phits for deroutes.

### 3.1.2 Polarized Routes

Polarized [5, 3] provides both minimal and non-minimal routes that are made hop-by-hop. Let $s$ be a source switch, $t$ the destination switch and $c$ the current switch $c$ in the route and $d(s, t)$ the distance from between the switch $s$ to the switch $t$. Polarized routes are constructed hop-by-hop ensuring that, in every hop, the following weight function is never decreased:

$$\mu_{s,t}(c) = d(c,s) - d(c,t). \quad (1)$$

Given a current switch $c$, any hop to a neighbour switch will either *approach*, *revolve*, or *depart* $s$ and $t$, resulting in nine possibilities. From these, in Table 1, are displayed just the 5 options allowed by Polarized routing. Entries of the table, $(\Delta s, \Delta t)^{\Delta \mu}$, provide the distance variations, respect to both source and destination, of making any allowed hop and their associated weight variation, which will be used for allocation prioritization. More formally, if $c_p$ is the neighbour switch of $c$ connected by the port $p$, then $\Delta s = d(s, c_p) - d(s, c)$ and $\Delta t = d(t, c_p) - d(t, c)$, satisfying $-1 \leq \Delta s, \Delta t \leq 1$. Similarly, it is defined $\Delta \mu_{s,t}(c) = \mu_{s,t}(c_p) - \mu_{s,t}(c) = \Delta s - \Delta t$, with a value $-2 \leq \Delta \mu_{s,t} \leq 2$. All the neighbours of the current switch $c$ with $\Delta \mu \geq 0$ will be candidates for a possible polarized route. To avoid cycles, candidates with $\Delta \mu = 0$ are filtered depending on whether they are closer to the source or to the destination. This may be managed by including a boolean field $d(c,s) < d(c,t)$ in the packet header and updating it with each move or by storing the distances in the routing table. Observe that all the information needed by Polarized is obtained by accessing twice (one indexed by $s$ and the other by $t$) to the routing tables, in which $\{-1, 0, 1\}$ values indicate whether the candidate port approaches, revolves or departs the considered switch.

|  | Approaches t | Revolves t | Departs t |
|---|---|---|---|
| **Departs s** | $(+1, -1)^2$ | $(+1, 0)^1$ | $(+1, +1)^0$ |
| **Revolves s** | $(0, -1)^1$ |  |  |
| **Approaches s** | $(-1, -1)^0$ |  |  |

Table 1: Distance variations to target and source, $(\Delta s, \Delta t)$, after a hop. Super index indicates $\Delta \mu$ value, which will be used in allocation decisions.

Finally, the candidates are given priorities depending on their $\Delta \mu$. The candidates in which $\Delta \mu$ is highest are given a no penalization of $P = 0$ phits. The candidates with a $\Delta \mu$ of 1 less are given $P = 64$ phits, and the ones with 2 less are given $P = 80$ phits.

Both Polarized and OmniWAR used a deadlock avoidance mechanism dependent of the network diameter, which employs the virtual channels at each port using the packet's hop count order [16, 25], referred to as a *ladder* in this context. Essentially, the $i$-th virtual channel is utilized when the packet has already passed through $i$ switch-to-switch links. The length of polarized routes in the HyperX are at most the double of the network diameter, resulting in the same number of required VCs as other similar related mechanisms to support adaptive non-minimal routing. On a fault-free 3D HyperX, a ladder of 6 VCs is required to obtain the highest performance. Nevertheless, when failures arise, the diameter of the network can grow, augmenting in consequence the length of the routes, and invalidating the use of a 6 step ladder. By using an escape subnetwork for avoiding deadlocks instead of a ladder, this restriction is lifted, allowing to select the best number of virtual channels according to performance and cost restrictions. The use of the rich polarized routes with an Up/Down based escape subnetwork allows SurePath to be used in the presence of faults with just 2 VCs, allowing to add more VCs for performance.

### 3.2 Opportunistic Up/Down Escape Routing Subnetwork

The escape subnetwork employed in this paper to avoid deadlock is based on AutoNet [31]. In its original form, any spanning tree suffices, with any path along the tree guaranteeing deadlock-free escape. The routing employed was called Up*/Down*, which means following the path in the spanning tree, which clearly implies using non-minimal routes. However, such subnetwork is insufficient in many circumstances, effectively replacing a deadlock into the marginal throughput of a tree. For example, in [14] a spanning tree escape subnetwork was used in Dragonfly networks, which could perform adequately under some conditions but could also provide insignificant throughput in others. With the aim of a robust network, in our proposal, a modification of this idea is made to construct the escape subnetwork. Instead of using a spanning tree, all the links of the network are used by adding shortcuts as in [35, 7]. Therefore, the escape subnetwork is enhanced by adding opportunistic shortcuts which allow having higher throughput in the subnetwork than only having up/down routes. This prevents performance degradation, and is one of the original contributions of our work.

Let us explain the construction of the escape subnetwork with support on Figure 2. Let us denote by $d(x, y)$ the graph distance from switch $x$ to switch $y$. At the beginning of the construction, an arbitrary switch is selected as root, which we denote by $r$. Then, the links of the topology are classified into two subclasses (Up/Down and horizontal) or colors (black and red). Let $(x, y)$ be the link joining nodes $x$ and $y$. Now, if $d(x, r) \neq d(y, r)$, the link is Up/Down or has black color. Otherwise, link $(x, y)$ is horizontal or has red color. For example, in Figure 2 it is shown this link classification for a $4 \times 4$ 2D HyperX. In the figure, the link joining switches $(1, 0)$ and $(1, 1)$ is black since $(1, 0)$ is at distance 1 from root $(0, 0)$, while $(1, 1)$ is at distance 2. On the contrary, the link joining switches $(1, 0)$ and $(2, 0)$ is red since their distance to root is the same.

Note that the set of black lines is not actually a tree, since a switch can have several parents. Nevertheless, these black links induce an Up/Down distance as follows. For example, switches $(1, 0)$ and $(2, 0)$ are at distance 2, following 1 link Up and another link Down. That is, the Up/Down distance is the minimum number of links traversed to reach from a switch $x$ to another $y$ starting with an Up subpath followed by a Down subpath. In each of these subpaths, every step get 1 closer/further to root $r$, respectively.

Now, red links will be used in an opportunistic way, that is, when they reduce the Up/Down distance between origin and destination, which provides adaptivity to the escape subnetwork. In Figure 2, let us consider that a packet has to go from switch $(0, 0)$ to switch $(1, 1)$. Since there are two possible Up/Down paths, the next step is selected between $(0, 1)$ and $(1, 0)$ by JSQ (joint to the shortest queue). Now, consider the same situation but from $(0, 1)$ and $(0, 3)$. In this case the Up/Down distance (using black links) is 2, but the direct red link from $(0, 1)$ to $(0, 3)$ reduces the Up/Down distance, so it is a preferable candidate. Note that the link from $(0, 1)$ to $(0, 2)$ is never considered as a candidate since it does not decrease the Up/Down distance.

A possible hardware implementation would consist of having a table at each switch $C$, indexable at every target switch $T$ and port $p$, giving the change in the Up/Down distance by traveling that link. Thus, for a given destination, the table gives a vector of ports, where each entry with a value greater than 0 representing a valid candidate in its associate escape channel.

In a high radix topology like HyperX, the black links of the almost-tree are few and should be avoided, lest congest the root. For this, it is important to use the horizontal links (or shortcuts), and we may use the available information to give them preference. Specifically we set $P = 112$ phit penalization for Up candidates and $P = 96$ phit penalization for Down candidates. Opportunistic links are given penalization of $P = 80, 64$ or $48$ phits, respectively, for candidates that reduce the Up/Down distance by 1, 2, or $\geq 3$. In the HyperX, minimal paths that use horizontal links reduce the Up/Down distance by $+2$ each step, so they are preferred. Thus, this escape subnetwork is actually able to use most minimal routes and can accept a reasonably high amount of load. Nevertheless, for most of the load, especially on adverse patterns, SurePath resorts to the routes provided by the routing algorithm. Next we evaluate performance of SurePath using either Omnidimensional or Polarized routings. As it will be seen, in both cases the performance is good, with some advantage of Polarized routing in an adversarial traffic pattern.

## 4 Experimental Setup

All the experiments carried out in this work have been made with the network simulator CAMINOS [4] using synthetic traffic patterns. The basic parameters employed in the simulations are detailed in Table 2. The topologies evaluated are a 2D HyperX of side 16 and a 3D HyperX of side 8 and their topological parameters are summarized in Table 3.

The routing mechanisms considered in the evaluation will be Minimal, Valiant [34], OmniWAR [24] and Polarized [5, 3]. OmniWAR routing has been designed *ad hoc* for HyperX networks. This routing allows for a number of deroutes in non-aligned dimensions of the network, as previously explained. For instance, if source and target switches are in the same row, OmniWAR does not allow routes outside that row. Both OmniWAR and Polarized are defined originally with Ladder VC management for deadlock avoidance. SurePath will be evaluated with two different sets of routes provided by these routings separately: Omnidimensional and Polarized. The notation for these configurations and the details of VCs usage are explained in Table 4. For the sake of a fair comparison, all routing mechanisms will be implemented using the same resources, that is, 4 in 2D HyperX and 6 in 3D HyperX ($2n$ for $n$-dimensional HyperX). In particular, Minimal uses these $2n$ VCs together with the restrictions provided by the ladder in two-by-two VC steps. Valiant, OmniWAR, and Polarized use the $2n$ VCs in a one-by-one VC step ladder. SurePath is configured to use the same VCs for fair comparison, although is possible to use less without performance degradation. It is important to note that the number of VCs strongly conditions switch cost and power consumption.

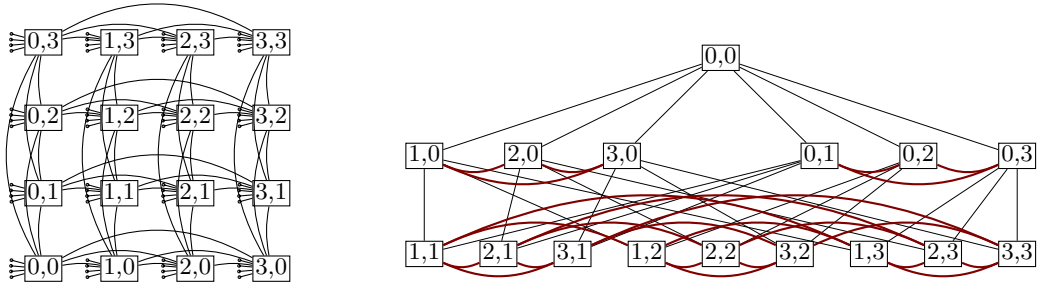

Figure 2: Left: A typical representation of a 4 × 4 HyperX network with 16 switches as rectangles with coordinates and 4 servers on each switch as dots. Right: Up/Down routing used for the escape subnetwork in the previous HyperX. There is always at least one Up/Down path, following the black straight lines. The other links (red curved lines) may be opportunistically employed when the Up/Down distance from the other endpoint to the target switch is lower than from the current switch.

| Parameter | Value |
| --- | --- |
| Input Buffer size | 8 packets |
| Output Buffer size | 4 packets |
| Flow control | Virtual cut-through |
| Packet length | 16 phits |
| Link latency | 1 cycle |
| Crossbar latency | 1 cycle (link) |
| Crossbar internal speedup | 2 |

Table 2: Simulation parameters.

| Parameter | 2D HyperX | 3D HyperX |
| --- | --- | --- |
| Switches | 256 | 512 |
| Radix | 46 | 29 |
| Servers per switch | 16 | 8 |
| Total servers | 4096 | 4096 |
| Links | 3840 | 5376 |
| Diameter | 2 | 3 |
| Avg. Distance | 1.8 | 2.625 |

Table 3: Topological parameters.

The synthetic traffic patterns considered in all the experiments are described next. Note that they are all *admissible*, which means that there is not endpoint contention.

In *Uniform* traffic pattern, each message is generated to a destination randomly chosen among the other servers. This is a classical benign pattern that may roughly represent general unstructured real traffic.

In *Random Server Permutation* traffic pattern, a random permutation of the servers is chosen. Then, the destinations of the messages are the result of applying the permutation to the sources of the messages. This traffic pattern could represent a situation in which every server pull a large file from another server, with those servers selected in a random but balanced way.

*Dimension Complement Reverse* was introduced in [24] for 3D HyperX. In this pattern, servers at a switch with coordinates $(x, y, z)$ send their traffic to other servers at the switch with coordinates $(\bar{z}, \bar{y}, \bar{x})$, where $\bar{x} = k-1-x$ is the complement modulo the side $k$. This is a pattern in which Valiant provides best throughput. In this work we adapt it also to 2D HyperX by using the coordinate of the server as another dimension. In the 2D variant, the server $(w, x, y)$ sends traffic to the server $(\bar{y}, \bar{x}, \bar{w})$, where $w$ is the coordinate of the server relative to the switch, and $(x, y)$ are the coordinates of the switch.

Importantly, in 3D HyperX we also evaluate the routing mechanisms under *Regular Permutation to Neighbour* traffic pattern. This is a new traffic pattern defined to emphasize the distinctions between Omnidimensional and Polarized sets of routes. In this traffic pattern, all servers attached to a switch will have destinations to servers attached to another switch, with this destination switch being the permutation of the source switch. This pattern is quite representative as a case where Minimal routing performs poorly, but a superior routing strategy can handle more load than what is accommodated using Valiant's load balancing scheme.

| Routing Mechanism | Routing Algorithm | VC management | Use of $2n$ VCs ($n = 2, 3$) | VCs required |
|---|---|---|---|---|
| Minimal | Shortest path | Ladder | 2 VCs for each step | $n$ |
| Valiant | Shortest path in each phase | Ladder | 1 VC for each step | $2n$ |
| OmniWAR | Omni-dimensional | Ladder | $n$ VCs minimal and $n$ VCs for deroutes | $2n$ |
| Polarized | Polarized | Ladder | 1 VC per step | $2n$ |
| OmniSP | Omnidimensional | SurePath | $2n-1$ VCs Pol + 1 VC Up/Down | 2 |
| PolSP | Polarized | SurePath | $2n-1$ VCs Pol + 1 VC Up/Down | 2 |

Table 4: Routing mechanisms evaluated.

Now, let us detail the switch permutation used for this Regular Permutation to Neighbour traffic pattern. For this traffic pattern, throughput is bounded by 0.5 when using aligned routes, as the ones provided by Omnidimensional. Let us assume a $k \times k \times k$ 3D HyperX or $K_k^3$ Hamming graph, with $k \geq 2$ even. Every $K_k^3$ is decomposed into $(k/2)^3$ hypercubes $K_2^3$ following the natural embedding. For example, the subgraph induced by switches with $\{0,1\}^3$ coordinates is a $K_2^3$ located at a corner of the HyperX. Over this hypercube it is built any directed Hamiltonian cycle of length 8. A permutation is then defined on this small hypercube such that every switch communicates to the next one in its Hamiltonian cycle of length 8. Now, this small permutation is extended to the whole network by repeating the same pattern in each of the $(k/2)^3$ hypercubes. The resulting traffic pattern ensures that in every $K_k$ subgraph (full rows in any dimension) there are exactly either 0 source/destination pairs or $k/2$ source/destination disjoint pairs.

This construction is shown in Figure 3 for $K_8^3$. The figure represents only one plane to show the traffic imbalance, since it is independent of the selected plane. The 16 paths visible on the figure are closed in cycles by using 2 vertical links each, to connect similar 3-length paths in its above plane. In a $8 \times 8 \times 8$ HyperX, as the one used for experimentation in this work, we will have $(8/2)^3 = 64$ small $K_2^3$ embedded hypercubes, each with its cyclic pattern.

Throughput for this traffic pattern can be calculated using network bisection. To this aim, consider a $K_k$ with $k/2$ source switches (a total of $k^2/2$ source servers) and $k/2$ switch destinations. Clearly, any route confined to this $K_k$ must contain at least one link connecting a source switch to a destination switch; and the total of these links is $k^2/4$. Thus, the flows of the $k^2/2$ servers have to go through those $k^2/4$ channels, which limits throughput to 0.5. Nevertheless, for any given link there are other $2(k-$

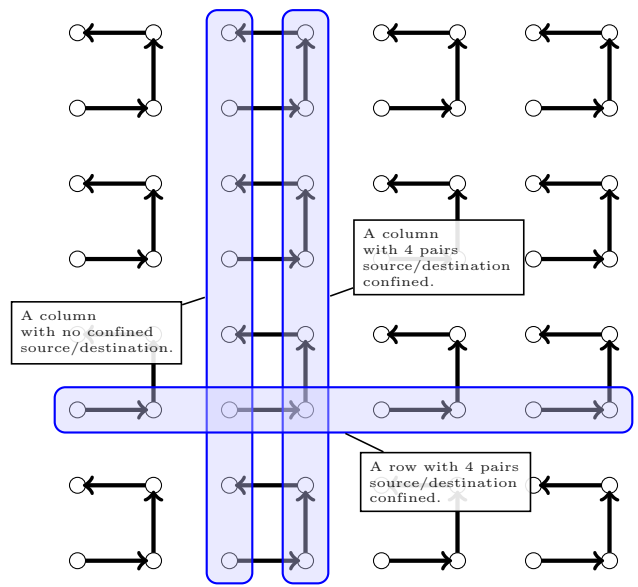

Figure 3: Source-to-destination pairs of the Regular Permutation to Neighbour in one plane of a 3D HyperX topology.

1) parallel ones, some of them uncongested, that may be used in a 3-hop route, from which Polarized routing takes advantage.

Three performance metrics are considered in the paper: average accepted throughput, average message latency and Jain index for load generation. First two are common metrics in the literature, and the third one is a metric that measures the fairness among the servers [18]. The Jain index follows this expression: $\frac{\left(\sum_{i=1}^{n} x_i\right)^2}{n \sum_{i=1}^{n} x_i^2}$, where $x_i$ is the load generated by server $i$ and $n$ is the total number of servers. A Jain index of 1.0 indicates perfect equity, with all servers injecting the same load. An index in the $0.98 - 1.00$ range, could be considered a good value. However, below 0.98 it entails certain unfairness, and the lower it is, the more disparity between servers.

# 5 Fault-free Performance Analysis

This section considers network performance in a fault-free scenario by SurePath routing mechanism. As explained before, this routing mechanism can use sets of routes provided by Omindimensional or Polarized. Hence, we evaluate both possibilities separately and compare it with other routings, including standard OmniWAR and Polarized routings. Results for 2D HyperX are shown in Figure 4 and 3D HyperX in Figure 5. In all cases the experiments show throughput, latency and Jain fairness index for the traffic patterns described before.

Let us first consider the 2D case, with the results in Figure 4. For Uniform traffic, all the routings, except Valiant, provide the same throughput. The same behaviour is observed for latency and Jain index. With the Random Server Permutation, OmniSP and PolSP provide the best results. Minimal routing has some problems due to saturation of some paths, as it can be appreciated in latency and Jain index results. Under Dimension Complement Reverse, Valiant achieves its theoretical 0.5, which is optimum, and matches the rest of routings, except Minimal. All routings except Valiant show Jain index drops. Note that escape subnetworks can generate this kind of effects, as considered in [11, 28]. In this particular case, both in Random Server Permutation and Dimension Complement Reverse, the distance between source and destination servers vary across the network. Therefore, there are differences in the accepted load between servers, since some servers have a more favourable destination than others, which results in Jain index deviations. As it can be observed, performance of OmniSP and PolSP routings are similar across the three traffic patterns.

In the 3D HyperX case, similar results are obtained for Uniform, Random Server Permutation and Dimension Complement Reverse traffic patterns. In all cases, SurePath implementations are the ones providing the highest throughput. Again, Jain index falls down for Surepath more notably in Dimension Complement Reverse. The difference between the Jain indexes of OmniWAR and Polarized comes from the different nature of the routes and the weight functions that are used for the VCs management. PolSP index drop is due to the escape network, as explained before. OmniSP has a more severe drop due to the accumulated effect of its routes and the escape network. In our opinion, this is a small trade off for tolerating faults without performance degradation and at a low cost, at it will be seen later.

However, Regular Permutation to Neighbour shows some differences between SurePath implementations. Under this traffic pattern, Minimal routing, as expected, has the worst performance. Polarized and PolSP routings provide the best performance, while OmniWAR and OmniSP can only equal Valiant performance. Note that in this traffic patterns it is necessary to get out of the $K_8$ subgraph (such as a row) containing both source and destination to allow a load higher than 0.5. Therefore, since Omnidimensional does not allow such deroutes, it cannot do it better. Therefore, these differences come from the different nature of the set of routes provided by Omnidimensional and Polarized. This traffic is of particular importance since it proves that the routes provided by the routing algorithm in the configuration of SurePath are important enough for obtaining a good performance.

# 6 Evaluating SurePath under Failures

In this section we consider the scenario of a HyperX network with removed links, representing failures. As stated before, it is important to note that a single fault can cause a routing algorithm or its deadlock avoidance mechanism not to work. For instance, DOR routing would leave switches disconnected when just a single link is removed. Another notable case is the Ladder VC management. It cannot work under many failures, as the number of hops a packet can do is upper bounded by the size of the Ladder, but packets could require to take longer routes in faulty scenarios. Under failures, SurePath routing mechanism keeps working thanks to its escape Up/Down subnetwork, which is constructed by means of a BFS, recomputing the routing tables, alike for Minimal routing.

In all the experiments both OmniSP and PolSP use 4 VCs (3 VCs for routing and 1 VC for escape subnetwork) since it is enough for maintaining outstanding performance in both healthy and faulty networks. This implies 33% cost reduction when compared to other mechanisms that use Ladder scheme in the case of 3D HyperX networks, which needs 6 VCs to work in fault-free conditions. It is notable that, regardless of its very low-cost, SurePath routing remains efficient and robust even in the presence of many failures.

In order to measure the performance of SurePath under failures, we consider two different scenarios which differ in the way faults happen. In the first scenario we consider a sequence of random faults. We begin with a healthy network, and then, the number of faults is increased in steps of 10 faults up to a total of 100. Since link failures are known to typically occur isolated [15] and mostly characterized by their mean time between failures, sets of random failures are a realistic model of common failures. In Figure 6 performance results of SurePath for 2D and 3D HyperX are shown. In the abscissa we have the number of faults and in ordinates accepted load. In both networks SurePath performance degrades smoothly, being most representative the degradation in Uniform traffic pattern, in which performance goes from 0.9 to 0.8. In the other traffic patterns the difference in negligible. Note that in both 2D and 3D cases, the sequence of faults include one increase of the

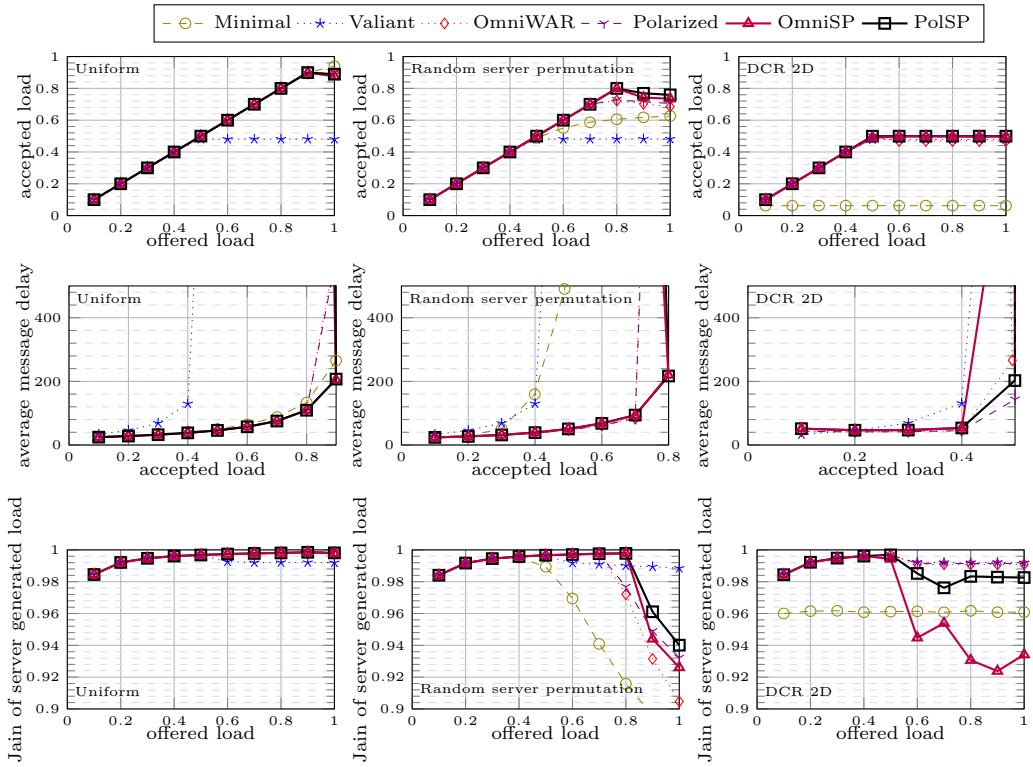

Figure 4: Performance for different routing mechanisms in 2D HyperX network under traffic: Uniform, Random Server Permutation and Dimension Complement Reverse.

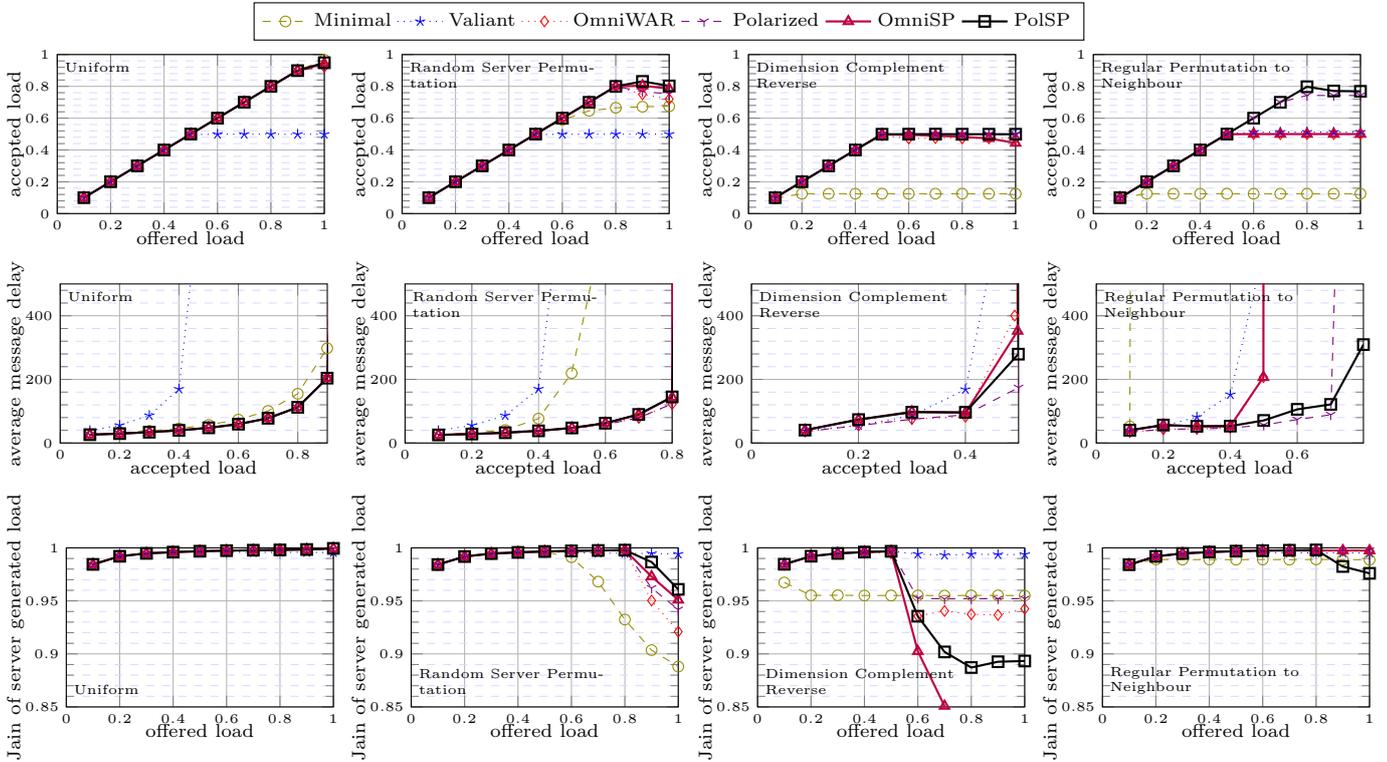

Figure 5: Performance for different routing mechanisms in 3D HyperX network under traffic: Uniform, Random Server Permutation, Dimension Complement Revers and Regular Permutation to Neighbour.

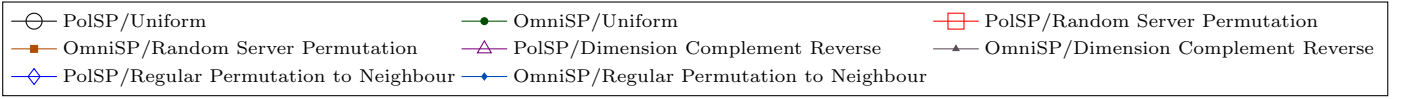

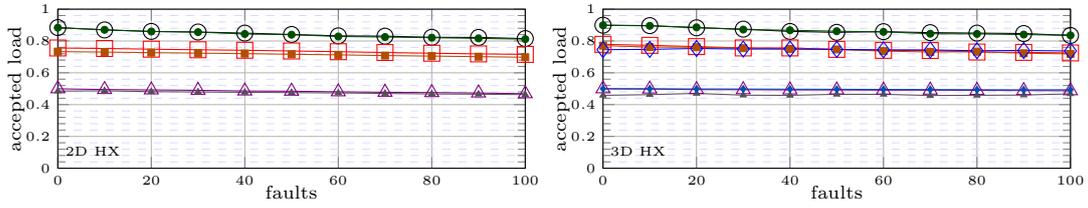

Figure 6: Throughput for successive failures from 0 to 100.

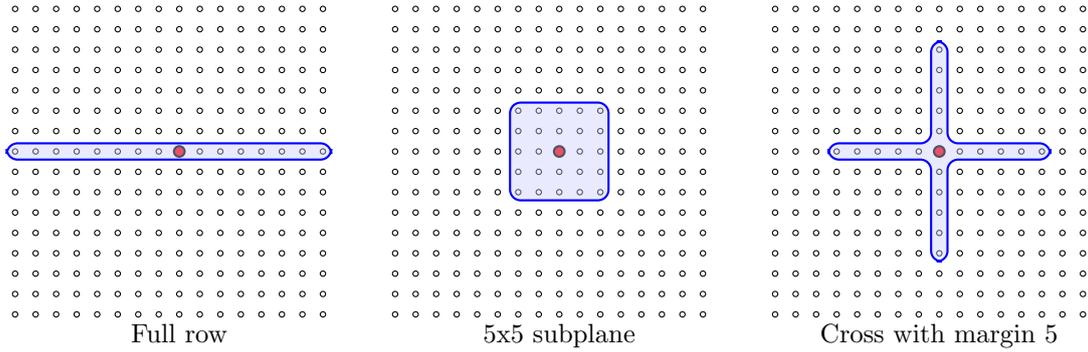

Full row     5x5 subplane     Cross with margin 5

Figure 7: The configurations of faults employed in 2D HyperX: Full Row, Cross and Subplane. The thick dot represents the switch selected as the root of the escape subnetwork.

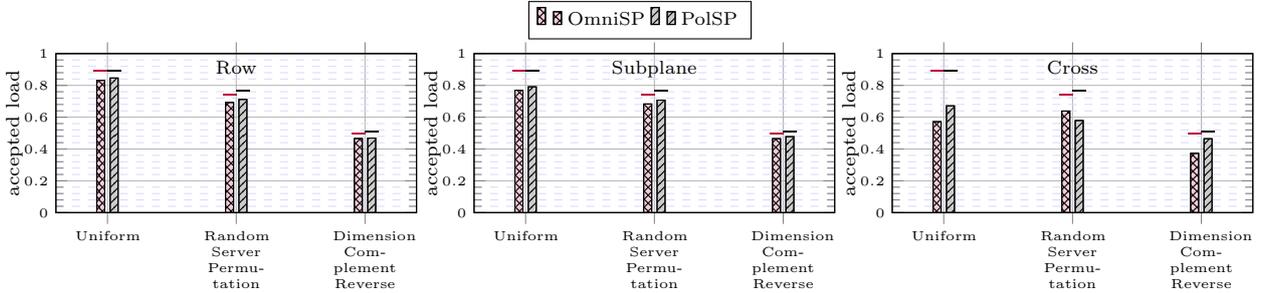

Figure 8: Throughput from 2D HX with all the edges contained in some shapes being removed. Top marks with the healthy network values have been added as reference.

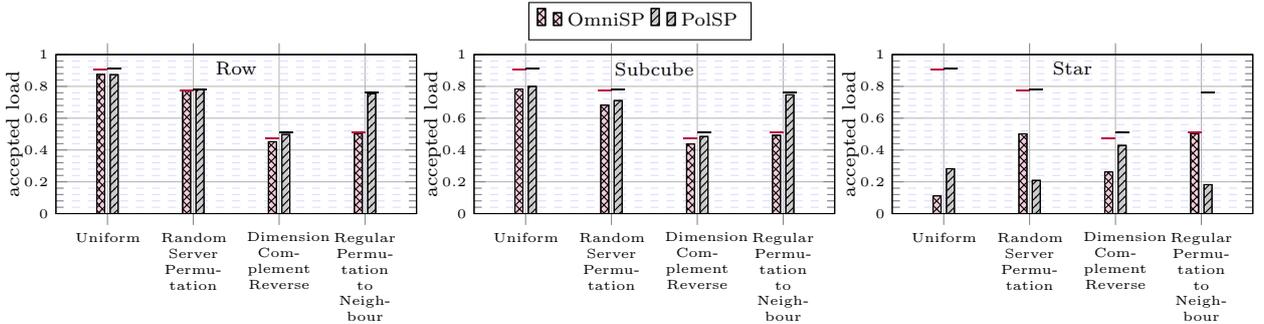

Figure 9: Throughput from 3D HyperX with all the edges contained in some shapes being removed. Top marks with the healthy network values have been added as reference.

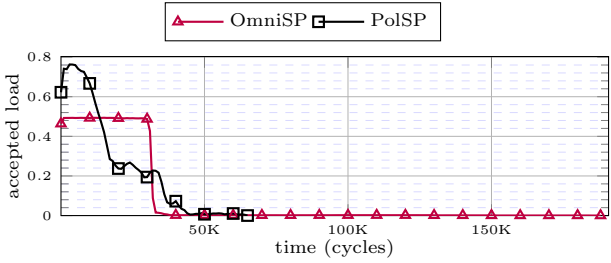

Figure 10: Completion Time for Regular Permutation to Neighbour Traffic Pattern in Star Fault Configuration.

diameter, which is not appreciable in performance. In particular, with this concrete set of faults the 2D HyperX has diameter 3 when the number of faults is 8 or greater. Also, the 3D HyperX has diameter 4 when the number of faults is 36 or greater.

In the second scenario, three particular fault configurations are examined: Row, Subplane and Cross. For the sake of clarity, in Figure 7 these configurations are graphically represented in the 2D HyperX evaluated. The 3D HyperX configurations follow the same idea. The Row configuration considers a row of switches (the ones highlighted in the figure) and the links that join two switches in the row are considered to have failed. This removes a $K_{16}$ subgraph with 120 edges of the complete network. The Subplane eliminates all the links corresponding to a square subset of switches, removing a $K_5^2$ subgraph with 100 edges. Finally, the Cross configuration does the same but with a row intersecting a column, while leaving a margin to prevent disconnecting its center. This cross removes the links in two $K_{11}$ subgraphs, which means a total of 110 links under fault. Importantly, all the configurations are designed such as the root of the escape subnetworks used by SurePath belongs to the set of switches under fault, seeking for a more stressful situation. These failure configurations are more artificial than those in the first scenario, but follow the *prepare for the unexpected* principle of reliability engineering [26].

In Figure 8, throughput results for OmniSP and PolSP in fault configurationss Row, Subplane and Cross are shown. These results are also considered under different traffic patterns: Uniform, Random Server Permutation and Dimensional Complement Reverse. Also, the lines above the bar graph indicate throughput in a healthy network, as a reference of performance degradation due to the presence of faults. As it can be seen, both OmniSP and PolSP provide almost the same throughput when a row of switches has failed, regardless the traffic pattern. Also, performance degradation of 11% indicates that, this fault configuration does not stress SurePath mechanism. Moreover, as it can be seen, the Subplane configuration provides similar results, which indicates that this configuration is also manageable. However, Cross configuration is clearly a challenge for the routing mechanism. In this case, Uniform traffic pattern is the one in which the difference between performance in a fault-free network or in the presence of faults is greater, with a 37% of performance drop. Note that this fault configuration is particularly adverse since it eliminates 2/3 of the links of the root of the escape subnetwork. Importantly, in these sets of experiments there is not a great difference coming from the sets of routes, that is, if we compare OmniSP versus PolSP.

Analogously, for 3D HyperX, similar experiments have been carried out. In this case we denote the fault configurations as Row, Subcube and Star, since they are 3D translations of the previous done for 2D. In *Row* it is considered a row with 28 faulty links. In the case of *Subcube*, a complete $K_3^3$ subcube with 81 links is eliminated, which is the 3D counterpart configuration to Subplane in the case of 2D. The *Star* configuration is the more extreme configuration considered in the paper, since it deletes a star with 63 edges which almost disconnects the root of the escape subnetwork, with only 3 remaining fault-free links.

The results in Figure 9 show that the situation is totally analogous to the 2D case in Row and Subcube configurations for Uniform, Random Server Permutation and Dimensional Complement Reverse traffic patterns. Differences arise when either the Regular Permutation to Neighbour traffic pattern or the Star faults configuration are involved. Next, we explain both.

The differences in Regular Permutation to Neighbour are originated by the different routes provided by Omnidimensional and Polarized routings, as illustrated in the experiment without faults. As it can be seen, in Row and Subcube fault configurations, PolSP provides better performance in Regular Permutation to Neighbour, proportional to the performance gains in a healthy 3D HyperX network. However, the results in the Star fault configuration are different. In this case, each of OmniSP and PolSP get ahead the other in different ways. Surprisingly, with these faults, OmniSP provides better throughput than PolSP in Regular Permutation to Neighbour, being able to perform nearly as in fault-free scenario. In the reverse, PolSP nearly provides the same throughput as with a healthy network for Dimension Complement Reverse pattern.

Let us analyze this particular case of the Permutation to Neighbour in a Start fault configuration with more detail. It is important to note that, as the root only has 3 alive links, there is an intrinsic in-cast contention. This is, the traffic generated at the 8 servers in the root can only go through these 3 links, and reciprocally, there are 8 neighbour servers that send traffic to the root and must go through the same 3 links in the opposite direction. The peak throughput shown in Figure 9 could hide these effects, so we show an additional experiment on Figure 10. In this experiment, each server in the network generates a load of 8000 phits, and the prominent figure is the *completion time*, where all packets are consumed. The figure shows the throughput obtained at each time of the simulation. In the first part of the graph, the throughput value is high, as it comprises the bulk of the load. After some time, roughly indicated by the already known peak throughput, most flows have finished and there is only a marginal throughput due to stragglers, including servers at the root. The last point in the figure corresponds to the application completion time. As it can be observed, although OmniSP provides more throughput in this situation, the completion time is $2.8\times$ the one experienced with PolSP.

The observed completion time can be explained satisfactorily. An ideal mechanism would accept the load into/from the root in about $1.33T$, being $T$ the number of cycles to accept the bulk of load with Omnidimensional. The value 1.33 comes from 8 servers in the root, over 3 links and by 0.5, which is the throughput by Omnidimensional in this traffic pattern. It is sensible to assume that traffic from the root and the rest to not overlap. Hence, the times add, since the 3 links at the root can be busy with external traffic. However, the routes provided by both Omnidimensional and the escape subnetwork can only use 1 of the 3 fault-free links in the root. This would mean that the completion time should be $5T$, which it is near the value in Figure 10.

The problem with Polarized routing is different. It is a routing eager to use many routes, which causes some problems with in-cast contention. The servers sending to the root can easily employ all of their many fault-free links. This may flood the network with packets that cannot cross the 3 links at the root, at a reasonable rate. This creates a congestion tree at the root that blocks other flows reducing peak throughput.

This extreme scenario has allowed us to determine different problems of both routings in relation to in-cast situations. Nevertheless, it must be highlighted that this scenario is more extreme than actually possible in practice. Furthermore, some of the issues can be addressed by avoiding to choose a switch with many faulty links as the root of the escape subnetwork.

## 7 Conclusions and Discussion

This paper presents and analyzes SurePath, a low-cost high performance fault-tolerant routing mechanism for HyperX networks. SurePath combines the set of routes provided by a routing mechanism plus an Up/Down escape subnetwork to avoid packet deadlocks. The routing mechanism not only gives high performance in fault-free HyperX network but also tolerates faults.

When configuring SurePath, it is important to start with a competitive routing algorithm. In the HyperX interconnection network we have selected the two best routing algorithms known: Omnidimensional and Polarized. Both routings are able to provide good enough routes to the mechanism. We have illustrated the impact of the routes when evaluating SurePath under Regular Permutation to Neighbour, a new traffic pattern that does not adapt well to Omnidimensional routes, resulting in a better performance with Polarized routes.

SurePath has proved its robustness when dealing with network failures. Maintaining its moderate cost, SurePath has revealed as an efficient and flexible solution for fault-tolerant applications. It has been shown how SurePath is able to obtain very high performance even in the presence of a large number of failures, both in a more realistic scenario (random faults) or in different stressful fault configurations. The experiments, which combine different traffic patterns and fault configurations, have shown that SurePath is able to maintain a good performance.

Finally, SurePath escape subnetwork is defined without any specific knowledge of the underlying topology, so it apparently could be used in any topology, not only in HyperX interconnection networks. However, it must be notice that HyperX networks can an have advantage because of their topological properties. In particular, in HyperX the escape subnetwork contains shortest paths or minimal routes. This is not true, for example, if the same mechanism would be used, as it is defined here, in Dragonfly networks. This means that more effort to adapt to other topologies should be done to attain the same level of performance. However, the results obtained for HyperX make SurePath a promising routing mechanism to be considered in other high-radix low-diameter interconnection networks.

## Acknowledgements

This work has been supported by the Spanish Ministry of Science and Innovation under contracts PID2019-105660RB-C22, TED2021-131176B-I00, and PID2022-136454NB-C21. C. Camarero is supported by the Spanish Ministry of Science and Innovation, Ramón y Cajal contract RYC2021-033959-I. R. Beivide is supported by The Barcelona Supercomputing Center (BSC) under contract CONSER02023011NG. Simulations were performed in the Altamira supercomputer, a node of the Spanish Supercomputing Network (RES).